\documentclass[aip,graphicx]{revtex4-1}
\usepackage[cp1251]{inputenc}
\usepackage[T2A]{fontenc}
\usepackage[english]{babel}
\usepackage{amssymb,latexsym,amsmath,amscd}
\usepackage{graphicx,color,framed}
\begin{document}
\title{A statistical field theory of salt solutions of 'hairy' dielectric particles}
\author{\firstname{Yury A.} \surname{Budkov}}
\email[]{ybudkov@hse.ru}
\affiliation{School of Applied Mathematics, Tikhonov Moscow Institute of Electronics and Mathematics, National Research University Higher School of Economics, Tallinskaya st. 34, 123458 Moscow, Russia}
\affiliation{Landau Institute for Theoretical Physics, Akademika Semenova av., 1-A, 142432 Chernogolovka, Russia}
\begin{abstract}
In this paper, we formulate a field-theoretical model of dilute salt solutions of electrically neutral spherical colloid particles. Each colloid particle consists of a 'central' charge that is situated at the center and compensating peripheral charges (grafted to it) that are fixed or fluctuating relative to the central charge. In the framework of the random phase approximation, we obtain a general expression for electrostatic free energy of solution and analyze it for different limiting cases. In the limit of infinite number of peripheral charges, when they can be modelled as a continual charged cloud, we obtain an asymptotic behavior of the electrostatic potential of a point-like test charge in a salt colloid solution at long distances, demonstrating the crossover from its monotonic decrease to damped oscillations with a certain wavelength. We show that the obtained crossover is determined by certain Fisher-Widom line. For the same limiting case, we obtain an analytical expression for the electrostatic free energy of a salt-free solution. In the case of nonzero salt concentration, we obtain analytical relations for the electrostatic free energy in two limiting regimes. Namely, when the ionic concentration is much higher than the colloid concentration and the effective size of charge cloud is much bigger than the screening lengths that are attributed to the salt ions and the central charges of colloid particles. The proposed theory could be useful for theoretical description of the phase behavior of salt solutions of metal-organic complexes and polymeric stars.
\end{abstract}
\maketitle
\section{Introduction}
Statistical mechanics of dielectric macromolecules is an emerging area of modern physical chemistry and condensed matter physics \cite{Budkov_review,Martin2016,Kumar_2014,Mahalik2016,Gordievskaya2018,Kumar_2009,Budkov2018,BudkovFPE2019,BudkovJML2019,BudkovEPJE1,BudkovEPJE2,Gurovich1994,Gurovich1995,
BudkovJCP2015,KolesnikovSoftMat2017,Dean_2012,Lu2015}. Nowadays, polarizable macromolecules, such as weak polyelectrolytes, polyampholytes, ionomers, blockcopolymers, polymerized ionic liquids, branched polymers, etc. have found a lot of applications in the modern industry of smart materials: from medicine and pharmacology to power and food industry. However, despite the great importance of these systems for industrial applications, the fundamental theory of dielectric macromolecules, based on the principles of statistical mechanics taking into account an internal electric charge distribution of macromolecules, at the moment remains at the initial stage of its development. Indeed, great efforts have been made by theorists to develop statistical models of polyelectrolyte solutions \cite{Dobrynin} and solutions of charged colloids \cite{Levin}, whose macromolecules carry nonzero net charge, localized on their surface, so that their thermodynamic properties are determined by the long-ranged Coulomb interactions. However, only several papers are devoted to the theoretical description of solutions and melts of electrically neutral macromolecules, having a complex internal electric charge distribution \cite{Martin2016,Gordievskaya2018,Budkov2018,BudkovFPE2019,BudkovJML2019,
Dean_2012,Lu2015}. The development of statistical theory, allowing us to calculate the equation of state and the static dielectric permittivity of solution on the basis of the microscopic model of inner charge distribution in the macromolecule is one of the most important problems of modern chemical engineering. One of the possible ways of solving this problem is to formulate a statistical theory of solution in terms of nonlocal field theory \cite{Efimov1967} and to apply the conventional field-theoretical methods, such as random phase approximation \cite{Borue1988}, variational approach \cite{Lue2006}, and self-consistent field theory \cite{Fredrickson}.

At the moment, several field-theoretical models of dielectric macromolecules taking into account their internal electric structure are presented in the literature. Several of them discuss the possible applications of field-theoretical models to description of the thermodynamic properties of solutions of dielectric macromolecules. In paper \cite{Martin2016} the authors formulated a field-theoretical approach to describing thermodynamic properties of solutions and melts of polarizable polymers with an arbitrary mechanism of chain flexibility. It is assumed that each monomeric unit is a spherical particle with a charge in its center and another opposite charge connected by a spring with the center in accordance with the standard Drude model. The authors claim that their general model can be applied to the description of a vast range of dielectric macromolecules, such as polyampholytes, polymeric ionic liquids, blockcopolymers, etc. In paper \cite{Budkov2018} the author of the present paper formulated the nonlocal field theory of dipolar molecules immersed in an electrolyte solution medium. In contrast to the earlier theories of polar fluids, describing polar molecules as point-like particles or hard spheres with a point-like dipole in their center \cite{Coalson1996,Abrashkin2007,Budkov2015,BudkovJCP2016,BudkovEA2018,McEldrew2018,Levin1999}, the proposed model describes the polar molecules as ionic pairs with a fluctuating distance between their charged centers. An arbitrary probability distribution function of distance between charged centers is attributed to each dipolar particle. In the framework of the random phase approximation, the author obtained a general expression for the electrostatic free energy of solution and a nonlinear integro-differential equation for the potential of self-consistent field, generalizing the Poisson-Boltzmann-Langevin equation, derived for the case of point-like polar molecules and ions \cite{Coalson1996,Abrashkin2007,Budkov2015}. Using the obtained self-consistent field equation, the potential of point-like test charge, situated in the media of solution of dipolar particles was derived in the linear approximation. It has been shown that for dipolar lengths corresponding to the protein molecules, the potential of a test charge strongly deviates from the Coulomb law (that is predicted within the local theory) at the distances order of several nanometers. In paper \cite{BudkovJML2019} the same author proposed a generalization of the previous nonlocal theory of solutions of dipolar particles, taking into account their chain association according to the 'head-to-tail' mechanism. Paper \cite{BudkovFPE2019} proposes a general nonlocal theory of solutions of electrically neutral soft molecules, described as a set of charged sites, interacting with each other through the Coulomb and arbitrary soft-core repulsive potentials that look like the Gaussian-core and Yukawa potentials. In the framework of the proposed theory within the random phase approximation, the author obtained a general expression for excess Helmholtz free energy of solution. As an illustration, the proposed general theory was applied to describing the phase behavior of the Gaussian-core dipolar model (GCDM), which is a direct extension of the well known Gaussian-core model (GCM) \cite{Likos} for the case of additional dipole-dipole interactions between the soft-core particles. An analytical expression has been obtained for the free energy of GCDM, generalizing the formerly obtained expression in \cite{Budkov2018} for taking into account the repulsive soft-core interactions between the dipolar particles and predicting the liquid-liquid phase separation of the solution with an upper critical point. It was pointed out that the obtained results could be used for analysing the phase behavior of aqueous solutions of protein macromolecules. Here, it is worth mentioning a recent paper \cite{Delaney2019}, where in the framework of the self-consistent field theory of polymer solutions and the random phase approximation the authors investigated liquid-liquid phase separation of solutions of intrinsically disordered proteins which are electrically neutral as a whole, but, at the same time, have a complex charge distribution and a soft-core structure. In both approaches, the authors theoretically predicted the binodal of liquid-liquid phase separation with an upper critical point for the salt-free solution, as well as in the presence of salt ions.

It is important to note that earlier the authors of paper \cite{Buyukdagli2013_JCP} developed the nonlocal field-theoretical model of the electrolyte solution with polarizable ions in accordance with the above mentioned Drude model and with an explicit account of the polar solvent, whose molecules are described by dimers of two oppositely charged centers, separated by a fixed distance. In paper \cite{Blossey2014_2} the authors theoretically investigated a thermodynamic behavior of electrolyte solution with explicit account of polar solvent that is under nano-confinement within the similar field-theoretical approach. In paper \cite{Buyukdagli2013_PRE} a nonlocal field theory of aqueous electrolyte solutions, taking into account higher multipole moments of the water molecules was proposed. However, we would like to point out that the aim of all these papers was mostly to find a microscopic justification of the {\sl nonlocal electrostatics}, formulated earlier -- the phenomenological theory, developed by Kornyshev and coauthors \cite{Kornyshev1980,Kornyshev1996,Bopp1996}, describing of solvation effects in low-molecular weight systems. However, in contrast to the above mentioned papers \cite{Budkov2018,BudkovJML2019,BudkovFPE2019}, papers \cite{Buyukdagli2013_JCP,Blossey2014_2,Buyukdagli2013_PRE} do not describe the thermodynamic properties of solutions of polarizable macromolecules, although they use an approach similar to the field-theoretical approach, formulated in \cite{Budkov2018}.\footnote{The main conceptual difference of the field theoretical approach that is formulated in \cite{Budkov2018,BudkovJML2019,BudkovFPE2019} from the one proposed in \cite{Buyukdagli2013_JCP,Blossey2014_2,Buyukdagli2013_PRE} consists in operating with an arbitrary distribution function $g(\bold{r})$ of the distance between charged centers of dipolar particles \cite{Budkov_reply}. Another difference is related to the fact that the field-theoretical approach in \cite{Budkov2018,BudkovJML2019,BudkovFPE2019} is based on the $NVT$ canonical ensemble, whereas in \cite{Buyukdagli2013_JCP,Blossey2014_2,Buyukdagli2013_PRE} -- on the $\mu VT$ grand canonical ensemble. The nonlocal statistical field theory for uniformly charged rods, based on the grand-canonical ensemble, was formulated in paper \cite{Lue2006}.}

As it was pointed out above, in paper \cite{Budkov2018} we proposed a general nonlocal statistical theory of salt solutions of dipolar particles, which could be used for describing thermodynamic properties of solutions of dipolar macromolecules, such as proteins and betaines. However, there is a wide class of macromolecules, whose internal electric structure cannot be reduced to two charged centers. The most natural generalization of the model of dipolar particle would be a model of an electrically neutral $"$star$"$ (see, fig. 1). In that case charge $q$ (further called the 'central charge') is placed in the geometrical center of a star and around it other 'peripheral' charges $q_{\alpha}$ ($\alpha=1,2,..,m$) are situated. The positions of peripheral charges relative to the central charge are fixed or fluctuating ones. We would like to note that such kind of charge configurations can be realized for polymeric stars \cite{Likos} and colloid particles \cite{Linse} in low-polar solvents with counterions, adsorbed on their surface. Another chemical systems, for which this model could be relevant are metal-organic complexes, whose molecules consist of multivalent metallic ion (central charge) and ligand counterions (peripheral charges), coordinated it through the organic spacers. To the best of our knowledge, up to now there have been no statistical models, describing thermodynamic properties of such kind of solutions. In the present paper we formulate such a theory.

The rest of the article is organized as follows. The second part formulates a general field-theoretic model of solutions of dielectric colloid particles. In the third part we consider a simplified case of identical peripheral charges on the colloid particles and derive an analytical expression for electrostatic free energy of solution, discuss the behavior of electrostatic potential of the test point-like charge in the salt solution of dielectric colloid particles with infinite number of peripheral charges. In the concluding forth part we discuss the main results of the paper and further prospects.

\section{Theory}
Let us consider a solution of $N_+$ point-like cations with charges $q_+ >0$, $N_-$ point-like anions with charges $q_- <0$, and $N_c$ electrically neutral colloid particles, confined in the volume $V$ at temperature $T$. We will model the solvent as a continuous dielectric medium with dielectric permittivity $\varepsilon$. As pointed out in the introduction, we assume that each colloid particle consists of central charge $q$ placed at the center of the particle and $m$ peripheral charges $q_{\alpha}$, $\alpha=1,2,..,m$, which are placed in the general case at fluctuating distances $\bold{\xi}_{\alpha}$ from the central charge, described by the probability distribution functions $g_{\alpha}(\bold{\xi}_{\alpha})$. The assumption of electrical neutrality of each colloid particle leads to the following condition for the charges $q+\sum_{\alpha=1}^{m}q_{\alpha}=0$. Since we consider only the case of a sufficiently dilute solution, we neglect all the intermolecular interactions except the Coulomb interactions.
\begin{figure}[h!]
\center{\includegraphics[width=1\linewidth]{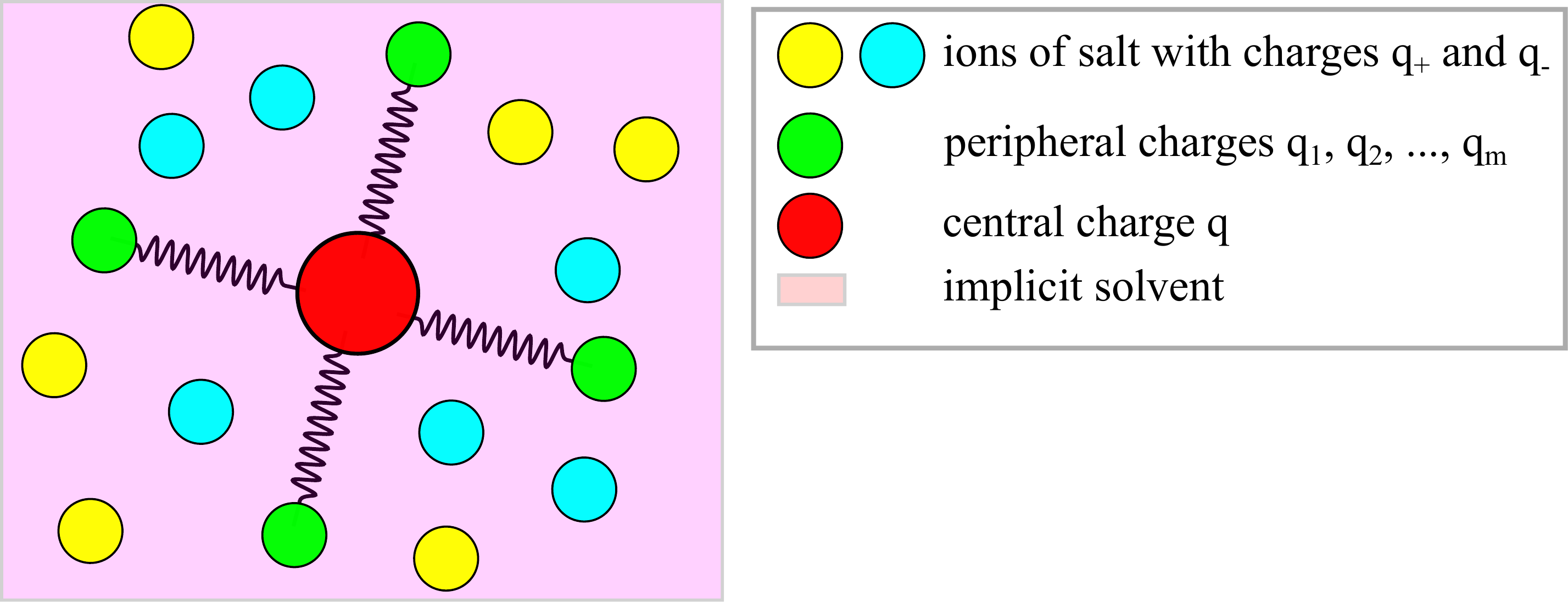}}
\caption{Schematic diagram of a colloid particle, surrounded by salt ions in a solvent medium.}
\label{fig1}
\end{figure}
Taking into account all the mentioned model assumptions, one can write the configuration integral of the system in the following form
\begin{equation}
Q=\int d\Gamma_s\int d\Gamma_{c}\exp\left[-\beta H\right],
\end{equation}
where
\begin{equation}
\int d\Gamma_c(\cdot)=\int..\int\prod\limits_{j=1}^{N_c} d\Gamma_{j}(\cdot)
\end{equation}
is the integration measure over internal configurations $\Gamma_{j}$ of the colloid particles, i.e.
\begin{equation}
\int d\Gamma_{j}(\cdot)=
\int\frac{d\bold{R}_j}{V}\int..\int\prod\limits_{\alpha=1}^{m}d\bold{\xi}_{j}^{(\alpha)}g_{\alpha}(\xi_{j}^{(\alpha)})(\cdot),
\end{equation}
where $\bold{\xi}_{j}^{(\alpha)}$ are the displacement vectors of peripheral charges relative to positions $\bold{R}_{j}$ of the central charges. The integration measure over the coordinates $\bold{r}_{k}^{(\pm)}$ of the salt ions is
\begin{equation}
\int d\Gamma_s (\cdot)=\int\prod\limits_{l=1}^{N_{+}}\frac{d\bold{r}_l^{(+)}}{V}\int\prod\limits_{k=1}^{N_{-}}\frac{d\bold{r}_k^{(-)}}{V}(\cdot),
\end{equation}
where the Hamiltonian of Coulomb interactions can be written as follows
\begin{equation}
\label{hamilt_el}
H=\frac{1}{2}\int d\bold{r}\int d\bold{r}'\hat\rho(\bold{r})G_0(\bold{r}-\bold{r}')\hat\rho(\bold{r}')=\frac{1}{2}\left(\hat\rho G_0 \hat\rho\right),
\end{equation}
where $G_0(\bold{r}-\bold{r'})=1/(\varepsilon|\bold{r}-\bold{r}'|)$ is the Green function of the Poisson equation and
\begin{equation}
\hat\rho(\bold r)=\hat\rho_{c}(\bold r)+\hat\rho_{i}(\bold r)+\rho_{ext}(\bold r)
\end{equation}
is the total microscopic charge density of the system;
\begin{equation}
\hat\rho_c(\bold r)=\sum\limits_{\alpha=1}^{m}q_{\alpha}\sum\limits_{j=1}^{N_c}
\left(\delta\left(\bold{r}-\bold{R}_{j}-\bold{\xi}_{j}^{(\alpha)}\right)-\delta\left(\bold{r}-\bold{R}_{j}\right)\right)
\end{equation}
is the microscopic charge density of the colloid particles;
\begin{equation}
\hat\rho_{i}(\bold{r})=q_{+}\sum_{k=1}^{N_{+}}\delta\left(\bold{r}-\bold{r}_k^{(+)}\right)+q_{-}\sum_{l=1}^{N_{-}}\delta\left(\bold r-\bold{r}_l^{(-)}\right)
\end{equation}
is the local charge density of the salt ions; $\rho_{ext}(\bold r)$ is the density of the external charges; $\beta = 1/k_{B}T$ is the inverse thermal energy and $k_{B}$ is the Boltzmann constant.

Further, using the standard Hubbard-Stratonovich transformation
\begin{equation}
\exp\left[-\frac{\beta}{2}(\hat\rho G_0\hat\rho)\right]=\int\frac{\mathcal{D}\varphi}{C}\exp\left[-\frac{\beta}{2}(\varphi G_0^{-1}\varphi)+i\beta(\hat\rho\varphi)\right],
\end{equation}
we arrive at the following expression for the configuration integral in the form of functional integral
\begin{equation}
Q=\int\frac{\mathcal{D}\varphi}{C}\exp\left[-\frac{\beta}{2}(\varphi G_0^{-1}\varphi)+i\beta(\rho_{ext}\varphi)\right]Q_c^{N_c}[\varphi]
Q_{+}^{N_{+}}[\varphi]Q_{-}^{N_{-}}[\varphi]
\end{equation}
with one-particle configuration integrals
\begin{equation}
Q_c[\varphi]=\int\frac{d\bold R}{V}\int d\sigma(\bold{\xi}_{1},..,\bold{\xi}_{m})\exp\left(i\beta \sum\limits_{\alpha=1}^{m}q_{\alpha}(\varphi(\bold{R}+\bold{\xi}_{\alpha})-\varphi(\bold{R}))\right),
\end{equation}
and
\begin{equation}
Q_{\pm}[\varphi]=\int\frac{d\bold r}{V}\exp(i\beta q_{\pm} \varphi(\bold r))
\end{equation}
with a normalized measure of integration over the displacements of peripheral charges
\begin{equation}
\int d\sigma(\bold{\xi}_{1},..,\bold{\xi}_{m})(\cdot)=\int d\bold{\xi}_{1}..\int d\bold{\xi}_{m}\prod\limits_{\alpha=1}^{m}g_{\alpha}(\bold{\xi}_{\alpha})(\cdot),
\end{equation}
i.e. $\int d\sigma =1$ and the following short-hand notations
\begin{equation}
(\varphi G_0^{-1}\varphi)=\int d\bold r\int d\bold r'\varphi(\bold r)G_0^{-1}(\bold r,\bold r')\varphi(\bold r'),~~(\hat\rho\varphi)=\int d\bold r \hat\rho(\bold r)\varphi(\bold r),
\end{equation}
\begin{equation}
C=\int \mathcal{D}\varphi\exp\left[-\frac{\beta}{2}(\varphi G_0^{-1}\varphi)\right].
\end{equation}
In the thermodynamic limit
$$V\to \infty,~N_{c,\pm}\to\infty, N_{c,\pm}/V\to n_{c,\pm}$$
we obtain \cite{Efimov1996,Budkov2018}
\begin{equation}
Q_c^{N_c}[\varphi]\simeq
\exp\left[n_c\int d\bold R\int d\sigma(\bold{\xi}_{1},..,\bold{\xi}_{m})\left(\exp\left(i\beta\sum\limits_{\alpha=1}^{m}q_{\alpha} (\varphi(\bold{R}+\bold{\xi}_{\alpha})-\varphi(\bold{R}))\right)-1\right)\right]
\end{equation}
and
\begin{equation}
Q_{\pm}^{N_{\pm}}[\varphi]\simeq \exp\left[n_{\pm}\int d\bold r(\exp(i\beta q_{\pm} \varphi(\bold r))-1)\right].
\end{equation}
Thus, we arrive at the following functional representation of the configuration integral
\begin{equation}
\label{func_int}
Q=\int\frac{\mathcal{D}\varphi}{C}\exp\left[-S[\varphi]\right],
\end{equation}
where the following functional
\begin{equation}
S[\varphi]=\frac{\beta}{2}(\varphi G_0^{-1}\varphi)-i\beta(\rho_{ext}\varphi)-W[\varphi]
\end{equation}
is introduced; the auxiliary functional $W[\varphi]$ can be written in the form
\begin{equation}
\nonumber
W[\varphi]=n_c\int d\bold R\int d\sigma(\bold{\xi}_{1},..,\bold{\xi}_{m})\left(e^{i\beta\sum\limits_{\alpha=1}^{m}q_{\alpha} (\varphi(\bold{R}+\bold{\xi}_{\alpha})-\varphi(\bold{R}))}-1\right)
\end{equation}
\begin{equation}
\label{funct}
+n_{+}\int d\bold r (e^{i\beta q_{+}\varphi(\bold r)}-1)+n_{-}\int d\bold r (e^{i\beta q_{-}\varphi(\bold r)}-1).
\end{equation}

Now, we obtain an expression for electrostatic free energy of solution in the framework of the random phase approximation \cite{Budkov2018}. Expanding functional $S[\varphi]$ in (\ref{funct}) into the power-law series in the vicinity of the mean-field configuration $\varphi^{(MF)}(\bold{r})=i\psi(\bold{r})$, satisfying the Euler-Lagrange equation
\begin{equation}
\label{EL_eq}
\frac{\delta S[\varphi]}{\delta\varphi(\bold{r})}\biggr\rvert_{\varphi=i\psi}=0,
\end{equation}
and truncating the series by the second order on $\varphi(\bold{r})$, we obtain
\begin{equation}
Q=\exp\left[-S[i\psi]\right]\int\frac{\mathcal{D}\varphi}{C}\exp\left[-\frac{\beta}{2}\left(\varphi \mathcal{G}^{-1}\varphi\right)+O[\varphi^3]\right],
\end{equation}
where we have introduced the following renormalized inverse Green function
\begin{equation}
\mathcal{G}^{-1}(\bold r,\bold r'|\psi)=k_{B}T\frac{\delta^2 S[i\psi]}{\delta \varphi(\bold{r})\delta \varphi(\bold{r}')}=G_0^{-1}(\bold r,\bold r')+S_{i}(\bold{r},\bold r')+S_{c}(\bold{r},\bold r')
\end{equation}
with the following short-hand notations
\begin{equation}
S_{i}(\bold{r},\bold r')=\beta\left(q_{+}^2n_{+}e^{-\beta q_{+}\psi(\bold{r})}+q_{-}^2n_{-}e^{-\beta q_{-}\psi(\bold{r})}\right)\delta(\bold r-\bold r'),
\end{equation}
\begin{equation}
\nonumber
S_{c}(\bold{r},\bold r')=\beta n_c\sum\limits_{\delta,\gamma}q_{\delta}q_{\gamma}
\left<e^{-\beta\sum\limits_{\alpha=1}^{m}q_{\alpha}\left(\psi(\bold{r}+\bold{\xi}_{\alpha}-\bold\xi_{\gamma})-\psi(\bold{r}-\bold\xi_{\gamma})\right)}
\left(\delta(\bold{r}-\bold r'+\bold{\xi}_{\delta}-\bold\xi_{\gamma})-\delta(\bold{r}-\bold r'-\bold\xi_{\gamma})\right)\right>_{\xi}
\end{equation}
\begin{equation}
\label{Sc}
-\beta n_c \sum\limits_{\delta,\gamma}q_{\delta}q_{\gamma}
\left<e^{-\beta\sum\limits_{\alpha=1}^{m}q_{\alpha}\left(\psi(\bold{r}+\bold{\xi}_{\alpha})-\psi(\bold{r})\right)}
\left(\delta(\bold{r}-\bold{r}^{\prime}+\bold{\xi}_{\delta})-\delta(\bold{r}-\bold{r}^{\prime})\right)\right>_{\xi}.
\end{equation}
The self-consistent field equation, obtained from the Euler-Lagrange equation (\ref{EL_eq}) has the following form
\begin{equation}
\label{SC_eq}
\Delta\psi(\bold{r})=-\frac{4\pi}{\varepsilon}
\left(\rho_{ext}(\bold{r})+\bar{\rho}_{i}(\bold{r})+\bar{\rho}_{c}(\bold{r})\right),
\end{equation}
where
\begin{equation}
\bar{\rho}_{i}(\bold{r})=q_{+}n_{+}e^{-\beta q_{+}\psi(\bold{r})}+q_{-}n_{-}e^{-\beta q_{-}\psi(\bold{r})}
\end{equation}
is the average charge density of the salt ions, whereas
\begin{equation}
\label{rho_c}
\bar{\rho}_{c}(\bold{r})=n_c \sum\limits_{\gamma=1}^{m}q_{\gamma}
\left<e^{-\beta\sum\limits_{\alpha=1}^{m}q_{\alpha}\left(\psi(\bold{r}+\bold{\xi}_{\alpha}-\bold\xi_{\gamma})-\psi(\bold{r}-\bold\xi_{\gamma})\right)}-e^{-\beta\sum\limits_{\alpha=1}^{m}q_{\alpha}\left(\psi(\bold{r}+\bold{\xi}_{\alpha})-\psi(\bold{r})\right)}\right>_{\xi}
\end{equation}
is the average charge density of the bound charge of colloid particles. Note that in the expressions (\ref{Sc}) and (\ref{rho_c}) we have introduced the following symbol
\begin{equation}
\left<(\cdot)\right>_{\xi}=\int d\sigma(\bold{\xi}_{1},..,\bold{\xi}_{m})(\cdot)
\end{equation}
for averaging over the configurations of peripheral charges. Therefore, formally calculating the Gaussian functional integral, we arrive at the following general expression for the configuration integral of solution in the random phase approximation \cite{Budkov2018,BudkovFPE2019}
\begin{equation}
\label{RPA_gen}
Q\approx \exp\left\{-S[i\psi]+\frac{1}{2}tr\left(\ln \mathcal{G} -\ln G_{0}\right)\right\},
\end{equation}
where symbol $tr(..)$ denotes the trace of the integral operator in accordance with the definition
\begin{equation}
tr(A)=\int d\bold{r}A(\bold{r},\bold{r}),
\end{equation}
where $A(\bold{r},\bold{r}^{\prime})$ is the kernel of operator $A$. In the absence of external charges (i.e., when $\rho_{ext}(\bold{r})=0$), the mean-field potential $\psi(\bold{r})=0$, and the mean-field contribution to the excess free energy of solution $F_{el}^{(MF)}=k_{B}T S[0]=0$. Therefore, in this case the excess free energy is determined by the thermal fluctuations of the electrostatic potential near the zero value. In that case the renormalized inverse Green function takes the simplified form
\begin{equation}
\nonumber
\mathcal{G}^{-1}(\bold r,\bold r'|0)=G^{-1}(\bold r,\bold r')=G_0^{-1}(\bold r,\bold r')+2\beta I e^2\delta(\bold r-\bold r')
\end{equation}
\begin{equation}
+\beta n_c \sum\limits_{\delta,\gamma}q_{\delta}q_{\gamma} \left<\delta(\bold{r}-\bold r'+\bold{\xi}_{\delta}-\bold\xi_{\gamma})-2\delta(\bold{r}-\bold r'-\bold\xi_{\gamma})+\delta(\bold{r}-\bold r')\right>_{\xi},
\end{equation}
while the excess free energy in the random phase approximation can be calculated by the following formula
\begin{equation}
\label{F_el}
F_{el}\approx \frac{k_{B}T}{2}tr\left(\ln G_0 - \ln G\right)=\frac{Vk_{B}T}{2}\int\frac{d\bold k}{(2\pi)^3}\ln\frac{G_0(\bold k)}{G(\bold k)},
\end{equation}
where $G_0(\bold k)=4\pi/(\varepsilon k^2)$ and $G(\bold k)=4\pi/(\varepsilon(k^2+\varkappa^2(\bold k)))$ are the Fourier-images of the Green functions, which become translation invariant functions in the thermodynamic limit. Thus, subtracting the electrostatic self-energy of particles from the final expression \cite{Borue1988}, we arrive at the following excess free energy of solution in the random phase approximation
\begin{equation}
\label{el_free_en_lin_2}
F_{el}=\frac{Vk_BT}{2}\int\frac{d\bold k}{(2\pi)^3}\left(\ln\left(1+\frac{\varkappa^2(\bold k)}{k^2}\right)-\frac{\varkappa^2(\bold k)}{k^2}\right)
\end{equation}
with the screening function \cite{Borue1988,Lue2006,Brilliantov1993,Khokhlov1982,Budkov2018}
\begin{equation}
\varkappa^2(\bold{k})=\kappa_{s}^2+\frac{4\pi n_c}{\varepsilon k_{B}T}\left(|q(\bold{k})|^2+\sum\limits_{\alpha=1}^{m}q_{\alpha}^2(1-|g_{\alpha}(\bold{k})|^2)\right),
\end{equation}
where the following auxiliary function
\begin{equation}
q(\bold{k})=\sum\limits_{\alpha=1}^{m}q_{\alpha}(1-g_{\alpha}(\bold{k}))
\end{equation}
and characteristic functions \cite{Budkov2018}
\begin{equation}
g_{\alpha}(\bold{k})=\int d\bold{r} g_{\alpha}(\bold{r}) e^{-i\bold{k}\bold{r}}
\end{equation}
are introduced.

Now let us derive a general expression for the electrostatic mean-field potential $\psi(\bold{r})$ of the external charge with the density $\rho_{ext}(\bold{r})$, immersed to the medium of the salt colloid solution in the regime of weak electrostatic interactions
\begin{equation}
\label{psi}
\psi(\bold{r})=\frac{4\pi}{\varepsilon}\int\frac{d\bold{k}}{(2\pi)^3}\frac{ \tilde{\rho}_{ext}(\bold{k})}{k^2+\varkappa^2(\bold{k})}=\int\frac{d\bold{k}}{(2\pi)^3}\frac{ 4\pi\tilde{\rho}_{ext}(\bold{k})}{\varepsilon(\bold{k})k^2},
\end{equation}
where
\begin{equation}
\tilde{\rho}_{ext}(\bold{k})=\int d\bold{r} \rho_{ext}(\bold{r})e^{-i\bold{k}\bold{r}}
\end{equation}
is the Fourier-image of charge density of external charges and
\begin{equation}
\label{diel_func}
\varepsilon(\bold{k})=\varepsilon\left(1+\frac{\varkappa^2(\bold{k})}{k^2}\right)
\end{equation}
is the dielectric function of the solution \cite{Kornyshev1980}.

Using the expression for the dielectric function, one can calculate the static dielectric permittivity of the salt-free colloid solution ($\kappa_s=0$) as follows
\begin{equation}
\varepsilon_b = \lim\limits_{k\to 0} \varepsilon(\bold{k}).
\end{equation}
Taking into account that at small $k$ values the characteristic functions can be expanded into the power-law series $g_{\alpha}(\bold{k})\simeq 1-k^2\left<\xi_{\alpha}^2\right>/2$, where $\left<\xi_{\alpha}^2\right>$ are the mean-square displacements of the peripheral charges with number $\alpha$. Thus, we arrive at the following expression
\begin{equation}
\varepsilon_b=\varepsilon + 4\pi \gamma_c n_{c},
\end{equation}
where
\begin{equation}
\gamma_c=\frac{\sum\limits_{\alpha=1}^{m}q_{\alpha}^{2}\left<\xi_{\alpha}^2\right>}{k_{B}T}
\end{equation}
is the polarizability of the colloid particle. For the case of dipolar colloid particles, when $m=1$ and $\left<\xi^2\right>=l^2/3$ ($l$ is the dipole length), we arrive at the earlier obtained expression \cite{Budkov2018}.

\section{The case of identical peripheral charges}
In this section we consider the case of identical peripheral charges $q_{\alpha}=-q/m$ with identical spherically symmetric distribution functions $g_{\alpha}(\bold{r})=g(|\bold{r}|)$. In this case the screening function takes the form
\begin{equation}
\label{varkappa}
\varkappa^2(\bold{k})=\kappa_{s}^2+\frac{4\pi q^2n_c (m+1)}{\varepsilon k_{B}Tm}(1-g(\bold{k}))\left(1-\frac{m-1}{m+1}g(\bold{k})\right),
\end{equation}
where $\kappa_s^2=8\pi I/(\varepsilon k_{B}T)$ and $I=(q_{+}^2n_{+}+q_{-}^2 n_{-})/2$ is the ionic strength of solution.

For the case of dipolar particles ($m=1$) expression (\ref{varkappa}) gives the previously obtained expression \cite{Budkov2018}
\begin{equation}
\varkappa^2(\bold{k})=\kappa_{s}^2+\frac{8\pi q^2n_c }{\varepsilon k_{B}T}(1-g(\bold{k})).
\end{equation}
Let us consider another limiting regime $m\gg 1$, when the peripheral charges can be described by continual charge clouds, surrounding the central charges. The screening function in that case takes the form
\begin{equation}
\label{screen_func}
\varkappa^2(\bold{k})\simeq\kappa_{s}^2+\frac{4\pi q^2n_c }{\varepsilon k_{B}T}(1-g(\bold{k}))^2.
\end{equation}

Note that at small $k$ values, when $g(\bold{k})\simeq 1-k^2 R^2$, the dielectric function (\ref{screen_func}) behaves as
\begin{equation}
\varepsilon(\bold{k})\simeq\varepsilon\left(1+L_{Q}^2 k^2+\frac{\kappa_s^2}{k^2}\right),
\end{equation}
where
\begin{equation}
L_{Q}^2=\frac{4\pi q^2n_c R^4}{\varepsilon k_{B}T}
\end{equation}
is the 'quadrupolar' length, introduced firstly as a phenomenological parameter in the paper of Slavchov \cite{Slavchov}; $R$ is the characteristic size of the charge cloud of the colloid particle. Note the theory, formulated here allows us to express the quadrupolar length through the microscopic parameters of the colloid particles. We would like to note that in the regime of weak electrostatic interactions in the presence of 'quadrupolar' particles in the solution the self-consistent field equation takes the following form
\begin{equation}
\label{Nonloc_eq}
\Delta \psi(\bold{r})-L_{Q}^2\Delta^2\psi(\bold{r})-\kappa_s^2\psi(\bold{r})=-\frac{4\pi}{\varepsilon}\rho_{ext}(\bold{r}).
\end{equation}

Below, we will analyze a behavior of electrostatic potential of a point-like test charge immersed in a salt solution of the quadrupolar colloid particles. Moreover, we will calculate an electrostatic contribution to the total free energy of this colloid solution in the absence of external charges.

\subsection{Potential of point-like charge}
As a simplest case, it is instructive to calculate the potential of point-like test charge $q_{0}$ in a salt solution of the quadrupolar colloid particles. Placing the test charge at the origin, taking into account that $\rho_{ext}(\bold{r})=q_{0}\delta(\bold{r})$ and using equation (\ref{Nonloc_eq}) in the Fourier representation, we obtain
\begin{equation}
\psi(\bold{r})\simeq \frac{4\pi q_0}{\varepsilon}\int\frac{d\bold{k}}{(2\pi)^3}\frac{e^{i\bold{k}\bold{r}}}{k^2+\kappa_s^2+L_{Q}^2k^4},
\end{equation}
which after the calculation of the integral yields the following regimes
\begin{equation}
\psi(\bold{r})=\frac{q_0}{\varepsilon r}\times
\begin{cases}
\frac{\exp\left(-\kappa_{-} r\right)-\exp\left(-\kappa_{+} r\right)}{\sqrt{1-4\kappa_s^2 L_{Q}^2}}, \kappa_s L_{Q}< \frac{1}{2}\,\\
\frac{r}{\sqrt{2}L_{Q}}\exp\left(-\frac{r}{\sqrt{2}L_Q}\right), \kappa_s L_{Q}= \frac{1}{2}\,\\
\frac{\exp\left(-\kappa_0 r\right)}{\sqrt{4\kappa_s^2 L_{Q}^2-1}}\sin\left(\frac{r}{\lambda}\right),\kappa_s L_{Q}> \frac{1}{2},
\end{cases}
\end{equation}
where $r=|\bold{r}|$ and the following short-hand notations
\begin{equation}
\kappa_{\pm}=\frac{\left(1\pm\sqrt{1-4\kappa_s^2L_{Q}^2}\right)^{1/2}}{\sqrt{2}L_{Q}},~~\kappa_0=\kappa_s\sqrt{1+1/(2\kappa_s L_{Q})},
\end{equation}
\begin{equation}
\lambda^{-1}=\kappa_s\sqrt{1-1/(2\kappa_s L_{Q})}
\end{equation}
have been introduced. Thus, we obtain that if the quadrupolar length is less than half of the Debye screening length of the salt ions, then the electrostatic potential monotonically decreases at long distances. However, when the quadrupolar length exceeds half of the screening length, the potential decreases with oscillations, characterized by wavelength $\lambda$. Note that such a behavior of the electrostatic potential of the point-like charge, surrounded by salt ions and quadrupolar particles, was first obtained in the paper of Slavchov \cite{Slavchov} within the pure phenomenological theory. We would like to stress also that the condition $\kappa_s L_{Q}= 1/2$ determines the equation for the Fisher-Widom line \cite{FisherWidom,Stopper2019} for the ionic component of the solution. It is also interesting to note that analogous oscillations of the electrostatic potential (the so-called overscreening effect) are observed in ionic liquids \cite{Bazant}.

\subsection{Electrostatic free energy of solution}
Now, let us analyze a behavior of the electrostatic free energy of the salt solution of quadrupolar particles for the Yukawa-type model distribution function \cite{Budkov2018,BudkovJML2019,BudkovFPE2019}
\begin{equation}
g(r)=\frac{1}{4\pi R^2r}\exp(-r/R).
\end{equation}
The latter has the following characteristic function
\begin{equation}
\label{char_func}
g(\bold{k})=\frac{1}{1+k^2R^2}.
\end{equation}
Using (\ref{el_free_en_lin_2}), one can obtain the analytical expression for electrostatic free energy only for the salt-free case ($\kappa_s =0$), i.e.
\begin{equation}
\label{Fex}
\frac{F_{el}}{Vk_{B}T}=-\frac{1}{6\pi R^3}\left((1+y_c)\sqrt{1+\frac{y_c}{4}}-\frac{9y_c}{8}-1\right),
\end{equation}
where $y_c=L_{Q}^2/R^2=4\pi q^2 n_c R^2/(\varepsilon k_{B}T)=\kappa_c^2 R^2$ and $\kappa_c=(4\pi q^2n_{c}/\varepsilon k_{B}T)^{1/2}$ is the inverse Debye screening length, attributed to the central charges of quadrupolar particles. It is instructive to analyze the electrostatic free energy (\ref{Fex}) of the salt-free solution of quadrupolar particles in two limiting regimes, namely
\begin{equation}
\frac{F_{el}}{Vk_BT}=
\begin{cases}
-\frac{5\pi q^4 R}{16 \left(\varepsilon k_{B}T\right)^2}n_{c}^2, y_c\ll 1\,\\
-\frac{\kappa_c^3}{12\pi},y_c\gg 1.
\end{cases}
\end{equation}
The first regime reflects the case, when size $R$ of the charged cloud is much less than the Debye length $\kappa_c^{-1}$. In this case, the electrostatic correlations of the colloid particles manifest themself as the effective van der Waals attraction -- the Kirkwood-Shumaker interactions \cite{Kirkwood1952,Avni2019}. This attractive interaction is related to the spatial fluctuations of the charged clouds of the colloid particles, placing at the sufficiently long distances from each other. In the opposite regime, when size $R$ of the charged cloud is much bigger than the Debye length $\kappa_c^{-1}$, the electrostatic free energy can be described by the Debye-Hueckel limiting law. In this limiting regime the colloid solution is described by the one component plasma (OCP) model \cite{BrilliantovOCP}. Indeed, in that regime the system is just a set of central charges $q$, immersed to the compensating homogeneously charged background of the overlapping charged clouds.

Let us calculate the electrostatic free energy of the salt solution, when the colloid particles concentration is much less than the salt concentration. In other words, we consider the case $y_{c}\ll 1$. The result takes the following form
\begin{equation}
\label{small_yc}
\frac{F_{el}}{V k_{B}T}=-\frac{\kappa_s^3}{12\pi}-\frac{\pi q^2 n_c}{\varepsilon k_{B}TR}h_{1}(y_s)-\frac{\pi q^4 Rn_c^2}{16(\varepsilon k_{B}T)^2}h_2(y_s)+O(n_c^3),
\end{equation}
where $y_s=\kappa_s^2 R^2$ and we have introduced the following auxiliary functions
\begin{equation}
h_1(y_s)=\frac{y_s(1+2\sqrt{y_s})}{(1+\sqrt{y_s})^2},
\end{equation}
\begin{equation}
h_2(y_s)=\frac{40y_s^{3/2}+8y_s^2+48y_s+25\sqrt{y_s}+5}{(1+\sqrt{y_s})^5}.
\end{equation}

The first term in (\ref{small_yc}) describes the well known Debye-Hueckel limiting law for excess free energy of diluted electrolyte solutions. The second term is the contribution of the electrostatic interactions between the colloid particles and salt ions. The third term describes the contribution of the colloid-colloid Kirkwood-Shumaker interaction, renormalized by electrostatic screening of salt ions. In the regime when $y_{s}\gg 1$ and $y_c\gg 1$ we have the following limiting relation
\begin{equation}
\label{DH_OCP}
\frac{F_{el}}{V k_{B}T}=-\frac{\kappa^3}{12\pi},
\end{equation}
where $\kappa=(\kappa_s^2+\kappa_c^2)^{1/2}$; the excess free energy (\ref{DH_OCP}) is the Debye-Hueckel limiting law for the free energy of OCP in the presence of salt ions.

\section{Concluding remarks and prospects}
In this paper, we have formulated a field-theoretical model of salt solutions of electrically neutral spherical colloid particles. In the framework of our model, each colloid particle consists of a central charge and peripheral compensating charges, grafted to the central charge and separated from it by fluctuating distances. This model is an extension of the previously formulated field-theoretical model of salt solution of dipolar particles \cite{Budkov2018}. In the framework of random phase approximation, we have derived a nonlocal self-consistent field equation with respect to the mean-field potential generated by fixed external charges and an expression for excess free energy of the colloid solution.

In the absence of external charges, we have obtained a general expression for the excess free energy of a solution. For the limiting case of infinite number of peripheral charges, when they form a continual spherically symmetric charged cloud around the central charge, we have estimated the electrostatic potential of the point-like test charge in a salt colloid solution. We have determined that at long distances from the point-like charge, the potential behaves as the potential of charge, surrounded by ions and quadrupolar particles, first discovered within the phenomenological theory by Slavchov \cite{Slavchov}. We have obtained the analytical expression for the quadrupolar length, first introduced as a phenomenological parameter by Slavchov in the same paper. We have shown that the qualitative behavior of the electrostatic potential at long distances depends on the ratio of the quadrupolar length to the screening length of ions. Namely, when the quadrupolar length is less than half of the Debye screening length of the salt ions, the electrostatic potential monotonically decreases at long distances. However, when the quadrupolar length exceeds half of the screening length, the potential decreases with oscillations, characterised by a certain wavelength. The condition of equality between the quadrupolar length and half of the screening length determines the Fisher-Widom line for this system. We would like to note that the structural transition, first discovered by Slavchov within the phenomenological theory and predicted within our microscopic theory, from a monotonic decrease of potential to its oscillating behavior at long distances is a new fundamental effect which needs to be confirmed by computer simulations and experiments in future.

For the salt-free solution of quadrupolar particles we have obtained an analytical relation for the excess free energy. We have shown that, if the Debye screening length, associated with the central charges of the quadrupolar particles is much bigger than the effective size of the charge cloud, then the electrostatic free energy is proportional to the square of concentration. In this case, the electrostatic correlations of the colloid particles are reduced to their effective pairwise van der Waals interactions -- the Kirkwood-Shumaker interactions \cite{Kirkwood1952}. In the opposite case, when the screening length is much smaller than the effective size of the charged cloud, the salt-free colloid solution can be described by the one-component plasma model \cite{BrilliantovOCP}. In the case of a nonzero salt concentration, we have obtained analytical relations for the excess free energy of solution, when the salt concentration is much higher than the concentration of colloid particles and for the case when the size of the charged cloud is much bigger than both the screening length of the salt ions and the screening length of the central charges. For the latter case, we have obtained a regime, when the solution is described by one-component plasma in the presence of salt ions.

In conclusion, we would like to note that in the present paper we have formulated a theoretical background for describing thermodynamic properties of salt solutions of colloid particles with a complex inner charge distribution. Such a theory could be relevant describing the phase behavior of solutions of complex colloid particles in low-polar media, where the counterions condense onto the colloid surface \cite{Levin1999,Linse}, so that their positions undergo thermal spatial fluctuations around the center of the colloid particle. Another example, where this theory could be applicable, is salt solutions of metal-organic complexes, consisting of multivalent metallic ion, coordinated by counterions through the organic spacers \cite{Xu_2014,Perez_2008}.  Note also that proposed in this paper theoretical background can be applied to describing the salt solutions of micellar aggregates and polymeric stars. However, these applications could be a subject of the future researches.

\begin{acknowledgements}
The model development was supported by the Russian Science Foundation project No. 18-71-10061. The results, presented in section III were funded by the RFBR according to the research project No 18-31-20015. The author thanks Nikolai Brilliantov and Alexei Victorov for fruitful discussions and valuable comments. The author also thanks anonymous Referee for valuable comments and suggestions.
\end{acknowledgements}

\end{document}